\documentclass[pre,aps,amsmath,amssymb,lengthcheck,superscriptaddress,floatfix]{revtex4-1}
\usepackage{graphicx}
\usepackage{dcolumn}
\usepackage{bm}
\usepackage[usenames,dvipsnames]{color}
\usepackage{psfrag}
\begin{document}

\title{Dynamics of the central-depleted-well regime in the open Bose-Hubbard trimer}

\author{Vittorio Penna}
\affiliation{
Dipartimento di Scienza Applicata e Tecnologia, 
Politecnico di Torino, Corso Duca degli Abruzzi 24, I-10129 Torino, Italy}
\affiliation{
CNISM, u.d.r., Politecnico di Torino, Corso Duca degli Abruzzi 24, I-10129 Torino, Italy
}
\date{\today}

\begin{abstract}
We study the quantum dynamics of the central-depleted-well (CDW) regime in a 
three-mode Bose Hubbard model subject to a confining parabolic potential.
By introducing a suitable set of momentum-like modes we identify the microscopic variables 
involved in the quantization process and the dynamical algebra of the model. We describe
the diagonalization procedure showing that the model reduces to a double oscillator. 
%
Interestingly, we find that the parameter-space domain where this scheme entails a discrete spectrum 
well reproduces the two regions where the classical trimer excludes unstable oscillations.
Spectral properties are examined in different limiting cases together with
various delocalization effects.
These are shown to characterize quantum states of the CDW regime in the proximity of 
the borderline with classically-unstable domains.
\end{abstract}


\maketitle

\section{Introduction}
\label{uno}
Small-size bosonic lattices have attracted considerable attention in the last decade since they 
allow the exploration of a rich variety of dynamical behaviors in which macroscopic nonlinear 
effects are triggered through a few controllable parameters \cite{Nemoto}-\cite{HDC}. 
Such systems, formed by small arrays of coupled condensates, are governed by the discrete
nonlinear Schr\"odinger equations 
\begin{equation}
i\hbar {\dot z}_j = U|z_j|^2 z_j - T\bigl ( z_{j+1} + z_{j-1} \bigr )
\label{dste}
\end{equation}
where $U$ is the boson interaction, $T$ represents the tunneling amplitude, 
and $|z_j|^2$ is the population of the condensate at well $j \in [1,M]$ in a $M$-well lattice. 
Adopted almost thirty years ago to model small molecular systems 
and study energy-localization 
effects \cite{ELS1}-\cite{KC1}, equations (\ref{dste}) 
still represent the basic theoretic model for studying 
the dynamics of solitons and of low-energy excitations in lattices \cite{FW}-\cite{flop} where 
bosons can experience both attractive ($U<0$) and repulsive ($U>0$) interactions.

The appealing feature of small-size lattices
is that equations (\ref{dste}) involve a number of dynamical variables which is small but sufficient
to make the system nonintegrable. Thus, while preserving
a character simple enough to allow a systematic analytic approach, mesoscopic lattices
display strong dynamical instabilities and a variety of behaviors 
(including chaos) typically occurring in longer chains. 
This circumstance has stimulated a considerable interest in revisiting
nonlinear behaviors \cite{CFFCSS}, \cite{WEHMC} first studied at the classical level
within the fully quantum environment of bosonic lattices \cite{Persist}-\cite{Alon}.
Quantum aspects become relevant for
lattices involving low numbers of bosons per well. 
In this case, a realistic description of microscopic processes is provided by
the second-quantized Bose-Hubbard (BH) model \cite{hald80}-\cite{jbcgz98}
\begin{equation}
H = \! \sum_i \! \left [ \frac{_U}{^2} n_i (n_i -1) - v_i n_i \right ]\! -T \! \sum_i (a_i a^+_{i+1} +h.c.),
\label{bhm}
\end{equation}
where operators $a_i$, $a^+_i$ obey commutators $[a_i, \! a^+_\ell] \!= \!\delta_{\ell i}$ and
$n_i = a^+_ia_i$ are number operators. Equations (\ref{dste}), the semiclassical counterpart of model (\ref{bhm}),
are easily recovered within both the coherent-state \cite{apPrl80,ejc}
and the continuous-variable \cite{bpvPra84} picture 
showing how operators $a_i$, $a^+_i$ are replaced by complex variables $z_i$, $z^*_i$.

Theoretical work on mesoscopic lattices has been mainly focused on the three-well array
(trimer) this being the simplest nonintegrable model of this class of systems. 
Many interesting aspects of {\it semiclassical} trimer have been explored such as
its unstable regimes with both repulsive \cite{FPpre67} 
and attractive \cite{joha1} interaction $U$, the emergence of chaos in the presence of 
parabolic confinement \cite{BFPprl90}, external fields \cite{Chong} or off-site interactions \cite{VF2},
and discrete breather-phonon collisions \cite{HDC}.

Almost in parallel, the impressive development of laser trapping techniques, 
has made concrete the possibility to engineer small-size arrays whose dynamics is
accessible to experiments. The most prominent example is the two-well system (dimer) \cite{Anker,Albiez}
obtained by superposing a (sinusoidal) optical potential on the parabolic potential trapping the condensate. 
The same scheme should enable the realization of linear chains with an arbitrary number of wells
by adjusting the laser wavelength (determining the interwell distance) and the parabolic amplitude.

In this scenario, the study of trimer dynamics provides a privileged standpoint
to better understand the quantum counterpart of nonlinearity and instabilities. 
This issue has been discussed in a series of papers examining the spectral properties
of {\it quantum} trimer \cite{Persist}-\cite{Kol}, its description within the phase-variable \cite{Mossman} 
and the Husimi-distribution \cite{TWK} pictures, and the inclusion of higher-order quantum
correlations within the multiconfigurational Hartree method
\cite{Alon}. Quantum trimer has been also used to model coherent transport with weak interaction \cite{SGS1}
and thermalization effects within the Fokker-Planck theory \cite{vardi}.
More recently, the single-depleted-well regime of the trimer has been studied 
in \cite{JJK} to evidence the quantum signature of oscillatory instabilities. 

In the same spirit, in this paper, we study the quantum aspects of the single-depleted-well 
regime for an open trimer trapped in a parabolic potential. Based on previous work \cite{BFPprl90}, 
\cite{jpa42}, we aim, in particular, to obtain 
1) a satisfactory quantum description of stable macroscopic oscillations characterizing this
special regime, and to detect 2) significant effects that distinguish the approach to unstable regimes.

The presence of the parabolic trap results in an effective local potential that favors
the occupation of the central well. Owing the absence of a closed
geometry, the translation-invariant single-depleted-well solution of ring trimers reduces, 
in the present contest, to a stationary solution where the central well is depleted and the
lateral condensates exhibit twin populations and coherent (opposite) phases. For this
reason the relevant regime will be called central-depleted-well (CDW) regime. 
Classically, this is represented by trajectories whose initial conditions are
close to the CDW fixed point. 

The most part of fixed points of the trimer dynamics, which have the form of collective-mode stationary
solutions, feature a strong dependence from interaction parameters. The change of the latter provides
various mechanisms whereby one can control the coalescence (or the formation) of fixed-point pairs and 
thus the onset of dramatic macroscopic effects \cite{jpa42}.
Different from the other fixed points, the CDW solution has the special feature
to be always present it being independent from the model parameters. This property is 
advantageous at the experimental level in that the conditions for realizing configurations close to the
CDW state appear to be weakly conditioned by the tuning of physical parameters.
The semiclassical study of the CDW state has shown how its dynamics displays 
both stable and unstable regimes. In general, trajectories close to the CDW state exhibit
oscillations both of the lateral macroscopically-occupied condensates and of the central condensate
which typically involves small fractions of the whole population.

The interest for the CDW regime is motivated by its complex, manifold character which,
in addition to a well-known, rich scenario of stable and unstable behaviors, features dynamical modes 
whose character is at the border between the classical and the quantum behavior. We will explore this
regime showing that an almost exact description can be achieved.

If the small fraction of the central condensate indeed corresponds to a few-boson population, 
the semiclassical picture must be replaced by a purely quantum-mechanical approach.
Apparently, the only variable that must be quantized is the order parameter of the central well while
those pertaining to lateral wells (with many bosons) are expected to maintain their classical form.
This possibility is only apparent. In this paper we develop a quantum picture based on replacing some of trimer 
space modes with collective, momentum-like modes. This alternative formulation of trimer dynamics and, in particular, 
of the CDW regime allows one to identify the dynamical variables that really feature 
a quantum behavior. The new description is particularly interesting in that the model we obtain can be diagonalized 
in an exact way by implementing the {\it dynamical algebra method} \cite{Zhang}.

In Section~\ref{due} we review the semiclassical trimer model and its quantum counterpart,
and propose our alternative description in terms of momentum-like modes.
Section~\ref{tre} is devoted to obtain the energy spectrum and the eigenstates
for the two regimes characterizing the CDW dynamics. The diagonalization is made possible
by recognizing that algebra sp(4) is the dynamical algebra of this model. 
In section~\ref{quattro} we discuss the energy-spectrum properties for various 
limiting cases and highlight the link between the parameter-space regions
where the spectrum has a discrete character and the regions corresponding
to classically stable oscillations. The occurrence of a continuous spectrum
is related to classical instability. 
Section~\ref{cinque} further illustrates this aspect by showing how various
delocalization effects characterize the approach to classically unstable regions. 
Section~\ref{sei} is devoted to concluding remarks.

\section{Trimer dynamics in the CDW regime}
\label{due}

For a three-well array including parabolic confinement equations (\ref{dste}) take the form
\begin{eqnarray}
i \dot z_j &=&  U |z_j|^2 z_j - v_j z_j - T \, z_2\, ,\,\,\, j=\,1,3
\label{zeq13}
\\
i \dot z_2 &=&  U |z_2|^2 z_2 - v_2 z_2  - T \, ( z_1 + z_3)
\label{zeq2}
\end{eqnarray}
where on-site potentials $v_j$ are such that $v_2 > v_1=v_3 $ reflecting the form of the external potential. 
With no loss of generality one can assume $v_2 = V$ and $v_1=v_3 =0$. The relevant semiclassical BH Hamiltonian
\begin{equation}
\label{sHam}
{\cal H} = \frac{U}{2} \sum^3_{j=1} |z_j|^4 -V |z_2|^2
-T \bigl[ z^*_2 (z_1 + z_{3} ) + C.C. \bigr ]\, ,
\end{equation}
where canonical variables $z_i$ obey the canonical Poisson brackets $\{ z_m^*, z_n \} = i \delta_{mn}/\hbar $,
exhibits a single constant of motion $N = \sum^3_{i=1} |z_i|^2$ ($\{ N , {\cal H} \}=0$) representing the total
boson number. Quantities $z_i$ and $|z_i|^2$ are interpreted, in fact, as the condensate order parameter 
and the boson population at the $i$th well, respectively, 
defined by $z_i = \langle a_i \rangle$ and $|z_i|^2 = \langle n_i \rangle$,
in the quantum-classical correspondence with the BH model \cite{jpa41}.

Among many interesting regimes, the CDW regime, formed by phase-space trajectories
whose initial condition are close to the CDW solution
\begin{equation}
\label{cdw}
z_2 =0 \, ,\quad  z_1 = {\sqrt \frac{N}{2} } \, e^{ i\varphi -iut/\hbar} = - z_3 \, , \,\,\, u = \frac{UN}{2}\, ,
\end{equation}
indeed represents a special case owing to its evident independence from interaction parameters $U$, $V$ and $T$. 
The application of linear-stability analysis to the CDW solution \cite{jpa42} 
shows that the relevant dynamics features both stable
and unstable subregimes depending on the value of $\tau = T/(UN)$ and $v= V/(UN)$. 
In particular, numerical simulations \cite{jpa42} show that the stable regime displays
a regular dynamics with periodic oscillations of the three populations possibly involving different time scales. 
For initial conditions close to the CDW configuration both the lateral and central populations exhibit small deviations
from solution (\ref{cdw}) thus confirming the fact that lateral condensates oscillate maintaining their macroscopic
character whereas the central well remains almost empty.
Similar to Hamiltonian (\ref{sHam}), its quantum counterpart 
\begin{equation}
\label{qHam}
{H} = \frac{_U}{^2} \sum^3_{j=1} n_j (n_j-1)  -V n_2
-T \bigl[ a^+_2 (a_1 + a_{3} ) + H.C. \bigr ]\, ,
\end{equation}
involves three space modes $a_j$. Only mode $a_2$, however, features
a true quantum character, modes $a_1$ and $a_3$ being related to macroscopic boson
populations $\langle n_i \rangle = \langle a^+_i a_i \rangle \simeq N/2$ for $i=1,3$. 
This reasoning is only apparently correct. By introducing the new set of dynamical variables
\begin{equation}
\label{ABC}
A = \frac{z_1 +z_3}{\sqrt 2}\, , \quad
B = \frac{z_1 -z_3}{\sqrt  2}\, , \quad
C = z_2 \, ,
\end{equation}
with nonzero canonical Poisson brackets $\{A^*, A \}= i /\hbar$, $\{B^*, B \}= i /\hbar$
we obtain the trimer equations of motion in the alternative form
\begin{eqnarray}
i \hbar \dot A &=&  \frac{U}{2} (|A|^2  + 2|B|^2) A + \frac{U}{2} B^2 A^* - {\sqrt 2} T \, C \, ,
\nonumber
\\
i \hbar {\dot C}  &=&  U |C|^2 C -V\, C\, -T {\sqrt 2} \, A \, ,
\label{eqABC}
\\
i \hbar \dot B &=&  \frac{U}{2} (|B|^2  + 2|A|^2) B + \frac{U}{2} A^2 B^* \, ,
\nonumber
\end{eqnarray}
equipped with the constant of motion $N= |A|^2 + |B|^2 + |C|^2$.
The ensuing version of the CDW solution is
$$
A=0 \, , \quad C=0\, ,\quad B= \, {\sqrt N } \, e^{ i\varphi -iut/\hbar}
\, ,\quad u= UN/2 \, ,
$$
showing that the entire boson population is attributed to mode $B$. In this scenario,
trajectories representing small deviations from the previous CDW state involve two ``microscopic" modes,
namely $A$ and $C$, whose populations are small with respect to the total boson number $N$.
The only macroscopic quantity is thus the population $|B|^2$ of mode $B$. Equations (\ref{eqABC}) 
makes it evident that the rapid phase oscillations of mode $B$ can be easily removed from the trimer dynamics.
By setting 
\begin{equation}
A = e^{i\varphi (t)}\, a
\, , \quad  
C = \, e^{i\varphi (t)} c \, , \quad 
B = e^{i\varphi (t)} b \, , 
\label{abc}
\end{equation}
with $b = {\sqrt N} + \xi$ and $\varphi (t)=\varphi-ut$, $\hbar \equiv 1$, where $a$, $\xi$ and $c$ 
describe small deviations from the CDW state,
the motion equations reduce to the linear form
\begin{equation}
\begin{cases}
& i {\dot a} \simeq u\, (a+ a^*) -T {\sqrt 2} \, c 
\cr
& {\-} \cr
& i {\dot c} \simeq  -(u+V) \, c \, -T {\sqrt 2} \, a   
\cr
& {\-} \cr
& i {\dot \xi} \simeq  u \, ( \, \xi + \xi^*  ) 
\cr
\end{cases}
\label{eqabc}
\end{equation}
if quadratic and cubic terms involving microscopic variables $a$, $\xi$ and $c$
are neglected. The effective dynamics of the CDW regime is thus driven 
by the first two equations involving modes $a$ and $c$, since $\xi + \xi^* =0 $ must
be imposed to ensure condition $N = {\rm const}$ in the time evolution of trimer.
Simple calculations show that the Hamiltonian 
corresponding to equations (\ref{eqabc}) reads
$$
{\cal H}_f = {\cal H} - u N = - \frac{UN^2}{4} +  \frac{UN}{4} \bigl (\xi + \xi^* \bigr )^2 
$$
\begin{equation}
+ \frac{UN}{4} \bigl (a + a^* \bigr )^2  - (V+u)\, |c|^2 - {\sqrt 2}T\, ( a^*  c + c^* a) .
\label{Hf}
\end{equation}
${\cal H}_f$ can be derived as well from Hamiltonian (\ref{sHam}) 
by substituting variables (\ref{abc}).
The time-dependent canonical transformations (\ref{abc})
giving $A$, $B$ and $C$ in terms of $a$, $\xi$ and $c$, also show that the canonical structure is preserved,
namely that $\{ X^*, X\} =  i /\hbar$ with $X = a,\xi ,c $.
%

\subsection{The quantum model and its dynamical algebra}
\label{due1}
The discussion of appendix \ref{app1} shows how, similar to its classical counterpart $\cal H$, 
quantum Hamiltonian (\ref{qHam}) can be reduced to the quadratic form
$$
H_f \simeq  - \frac{U}{4}N^2 + \frac{u}{2} (\xi +\xi^+)^2
+ u n_A + \frac{u}{2} \Bigl [ A^{+2} + A^2 \Bigr ]
$$
$$
-(V+u) n_C - {\sqrt 2} T ( A^+ C + C^+ A)\, .
$$
This is achieved by introducing quantum collective modes 
$A = (a_1+a_3)/ \sqrt 2$, $B = (a_1 -a_3)/ \sqrt 2$ and $C= a_2$ and implementing a suitable
time-dependent unitary transformation ${\cal U}_t$ able to eliminate the macroscopic dynamics of 
mode $B$ described, at the classical level, 
by factor $e^{i\varphi (t)}$ in equations (\ref{abc}). The crucial point is that in the
new scenario quantum modes $A$, $\xi = B- {\sqrt N}$ and $C$  
play the same role of classical
modes $a$, $\xi$, $c$ in equations (\ref{eqabc}).
%
Operators $A$, $\xi$ and $C$ satisfy the standard commutation relation $[A, A^+] = 1$, $[C, C^+] = 1$
and $[\xi, \xi^+] = 1$. The interesting part of $H_f$ is 
$$
H_0 \simeq 
\frac{u}{2} \Bigl [ A^{+2} + A^2 +2n_A\Bigr ] -(V+u) n_C - {\sqrt 2} T ( A^+ C + C^+ A),
$$
which should characterize the quantum dynamical behavior of CDW-like states.
$H_0$ exhibits an unusual, composite form where, in addition to the standard coupling $A^+ C + C^+ A$ 
between bosonic modes, mode $A$ involves a pair of two-boson creation/destruction terms, $A^{+2}$ and $A^2$,
typically causing squeezing effects in atom-photon interaction models of quantum optics. 
These are also responsible for considerably increasing the complexity of the dynamical algebra of $H_0$.

One should recall that, given a Hamiltonian $H_0$, its dynamical algebra is the set of operators $D_i$ 
forming an algebraic structure, with definite commutators $[D_r, D_s] = i f_{rsh} D_h$,
based on which $H_0$ can be expressed as a hermitian linear combination $H_0 = \sum_i \nu_i D_i$.
In the absence of $A^{+2}$ and $A^2$ the dynamical algebra of $H_0$ is the spin algebra su(2) formed by generators
$$
D_+ = A^+ C \, ,\,\,\, D_- = C^+ A \, ,\,\,\, D_3 = (n_A -n_C)/2
$$
with $[D_3, D_\pm ] = \pm D_\pm $ and $[D_+, D_- ] = 2 D_3$,
in the well-known two-boson Schwinger realization. In this case, the diagonalization
of $H_0 = \nu_3 D_3 + \nu D_+ + \nu^* D_-$ with $\nu_3 \in {\mathbb R}$, $\nu \in {\mathbb C}$ 
would reduce to perform a simple rotation $S \in$ SU(2) such that 
$S H_0 S^+ =  {\sqrt{ \nu_3^2 + |\nu|^2}}\, D_3$ where $D_3$ is, by definition, 
the diagonal generator of su(2).

Owing to $A^{+2}$ and $A^2$ the dynamical algebra is the more complex symplectic algebra sp(4),
reviewed in appendix \ref{app2}, involving ten independent generators. To simplify the diagonalization
of $H_0$ it is advantageous to rewrite it in terms of canonical operators $x$, $y$, $q$ and $p$
such that $[x,y]= i$, $[q,p]= i$ defined by
\begin{equation}
\begin{cases}
& x= {(A+A^+)}/{\sqrt 2}\, ,\,\,\, y= -i {(A-A^+)}/{\sqrt 2}\, ,
\cr
& {\-} \cr
& q= {(B+B^+)}/{\sqrt 2}\, ,\,\,\, p= -i {(B-B^+)}/{\sqrt 2}\, .
\cr
\end{cases}
\label{xyqp}
\end{equation}
The final form  of $H_0$ is
\begin{equation}
H_0 = u \left  [ x^2 - \frac{1+2v}{2} ( q^2+p^2)
- 2{\sqrt 2} \tau (xq+yp) \right ]
\label{H0}
\end{equation}
where $u = UN/2$, $v = V/(UN)$ and $\tau = T/(UN)$. Even if we have assumed $V >0$,
reflecting the form of the parabolic-potential profile, we shall consider as well  
the possibility to realize negative $V$, representing the presence of a repulsive
central potential. In the latter case the three local potentials $v_j$ mimic the
mexican-hat profile.
The critical value where significant changes are expected
is $v =-1/2$, below which $-(1+2v)/2$ becomes positive.


\section{Energy spectrum}
\label{tre}

Hamiltonian $H_0$ can be diagonalized by a many-step process where
one exploits the knowledge of the transformations of group Sp(4) and,
in particular, of the effects of their action on momentum and coordinate operators.
Diagonalization is achieved by combining in a suitable way 
squeezing transformations $S_\alpha$, standard rotations $U_\theta$ and hyperbolic 
transformation $D_\phi$.
The latter belong to group Sp(4) they being generated, through the usual Lie-group 
exponential map, by generators $Q_1$, $J_2$ and $Q_3$,
respectively, of sp(4) (see Appendix \ref{app2}). 

\subsection{Diagonalization of case $v> -1/2$}
\label{3A}

\noindent
In this case $H_0$ can be diagonalized by a three-step process where
the diagonal Hamiltonian can be shown to be given by
$H_3 = W \, H_0 W^+$ with $W= D_\phi U_\theta S_\alpha$.
\medskip

\noindent
{\it First step}. $S_\alpha  = e^{-i \alpha Q_1  } = e^{i \alpha (xy-qp)/2 }$
\begin{equation}
\begin{cases}
& S_\alpha x S^+_\alpha = xe^{+\alpha}, \,\, S_\alpha y S^+_\alpha = ye^{-\alpha},
\cr
& {\-} \cr
& S_\alpha q S^+_\alpha = xe^{-\alpha}, \,\, S_\alpha p S^+_\alpha = ye^{+\alpha},
\cr
\end{cases}
\label{Ualfa}
\end{equation}
The action of $S_\alpha$ on Hamiltonian $H_0 $ allows one to
get the same coefficient (up to a factor $-1$) for $x^2$ and $q^2$
$$
H_1 = S_\alpha H_0 S_\alpha^+ = u {\sqrt{s}}  \left [ x^2 - q^2 - s p^2 
- \sigma (xq\, + yp ) \right ]
$$
$$
\sigma \equiv { 2{\sqrt 2}\, \tau}/{ {\sqrt{s}} } \, ,\quad s = v +1/2 \, .
$$
provided the condition  $e^{2\alpha} = \sqrt{s}$ defining $\alpha$ is imposed. 
\medskip

\noindent
{\it Second step}.
$U_\theta  = e^{-2i \theta J_2  } = e^{-i \theta (xp-qy) }$
\begin{equation}
\begin{cases}
& U_\theta x U^+_\theta = x C_\theta + q S_\theta, \,\, U_\theta y U^+_\theta = y C_\theta+ p S_\theta,
\cr
& {\-} \cr
& U_\theta q U^+_\theta = q C_\theta - x S_\theta, \,\, U_\theta p U^+_\theta = p C_\theta - y S_\theta,
\cr
\end{cases}
\label{Utheta}
\end{equation}
where $C_\theta = \cos \theta $ and $S_\theta = \sin \theta$. Transformation $U_\theta$ generates
the new Hamiltonian $H_2 = U_\theta H_1 U_\theta^+$ introducing an undefined parameter $\theta$
whereby terms depending on $xq$ can be suppressed. This condition is achieved
by imposing $2 \sin (2 \theta) - \sigma \cos (2 \theta)  = 0$ which provides a complete
definition of $\theta$ through ${\rm tg} (2\theta) = \sigma/2$. As a consequence, the new Hamiltonian takes the form 
\begin{eqnarray}
H_2 &=& \!\! \frac{u\, \sqrt{ s } }{2 \sqrt {4 + \sigma^2}} \Bigl [ ( 4 + \sigma^2 ) \, ( x^2-q^2) 
- ( 2s + \sigma^2 )  (p^2 - y^2)
\nonumber \\ 
\nonumber \\
&-& s \, {\sqrt {4 + \sigma^2 }}\, (p^2 + y^2)  \,
+ \, 2( v -3/2 )\, \sigma \, yp   \, \Bigr ]\, .
\nonumber
\end{eqnarray}
%
%
\medskip

\noindent
{\it Third step}.
$D_\phi  = e^{2i \phi Q_3  } = e^{i \phi (xp + qy) }$
\begin{equation}
\begin{cases}
& D_\phi x D^+_\phi = x c_\phi + q s_\phi, \,\, D_\phi y D^+_\phi = y c_\phi - p s_\phi,
\cr
& {\-} \cr
& D_\phi q D^+_\phi = q c_\phi + x s_\phi, \,\, D_\phi p D^+_\phi = p c_\phi -y s_\phi \, ,
\cr
\end{cases}
\label{Uphi}
\end{equation}
with $c_\theta = {\rm ch} \phi $ and $s_\theta = {\rm sh} \phi$.
This hyperbolic transformation has the property to leave $x^2-q^2$ and $y^2-p^2$ unchanged.
By acting on the last two terms of $H_2$ one can exploit parameter $\phi$ to eliminate term $yp$
in the final Hamiltonian. This is given by $H_3 = D_\phi H_2 D^+_\phi$ which reduces to a
linear combination of $x^2$, $q^2$, $y^2$ and $p^2$ if condition
$$
2 {\rm sh} (2 \phi)\,  (v +1/2)\, {\sqrt {4 + \sigma^2}} +  (2v -3)\, \sigma  \, {\rm ch} (2 \phi)=0 
$$
is satisfied. The latter condition gives the formula
\begin{equation}
{\rm th} (2 \phi)\,  = \, -  \frac{(2v -3)\, \sigma }{(2v +1)\, {\sqrt {4 + \sigma^2}} }
\label{th1}
\end{equation}
whereby $s_\theta$ and $c_\theta$ can be expressed in terms of the interaction parameters.
The final Hamiltonian reads
\begin{eqnarray}
H_3 
&=& \frac{u\, \sqrt{ s } }{2 \sqrt {4 + \sigma^2}} \, \Bigl [ ( 4 + \sigma^2 ) \, x^2 \, - ( 4 + \sigma^2 ) \, q^2 
\nonumber \\ 
\nonumber \\
&-& \, \left ( 2s + \sigma^2 + \frac{\Delta}{2} \right ) \, p^2 \, 
+ \, \left  ( 2s + \sigma^2 - \frac{\Delta}{2} \right ) \, y^2 \, 
\Bigr ]
\nonumber
\end{eqnarray}
where
$$
\Delta = \,\frac{ 4  }{\sqrt{ v +1/2 }}\,  {\sqrt { (v +1/2)^3\, +8\tau^2 \, ( v -1/2 ) }} \, .
$$ 
In order to ensure
that $\Delta$ assumes real values, the latter formula requires that $(v +1/2)^3\, +8\tau^2 \, ( v -1/2 )  > 0$.
This condition, combined with $v > -1/2$ gives
\begin{equation}
\tau \, <  f(v) \equiv {\sqrt {\frac{(v +1/2)^3}{8 (1/2 -v) } }}\quad {\rm for} \quad -1/2 < v \, .
\label{area1}
\end{equation}
which identifies a well defined portion ${\cal D}_+$ of the $v\tau$ parameter space.
Curve $\tau = f(v)$, denoted by $\Gamma_+$, is represented by the red curve of Fig. 1.
Hence, among the four domains visible in Fig. 1, ${\cal D}_+$ corresponds to the upper region 
bounded by $\Gamma_+$ from below. For $(v, \tau) \in {\cal D}_+$ the present diagonalization scheme
assigns a well-defined discrete spectrum to Hamiltonian $H_0$. Note that, for $v \ge 1/2$,
no upper limit constraints the range of $\tau$.
A second domain exhibiting a discrete spectrum can be found for $v \le -1/2$ which will
be identified in the sequel. 

The energy spectrum is easily worked out by observing that
Hamiltonian $H_3$ can be rewritten in terms of two harmonic-oscillator Hamiltonians
$$
H_3= 
\frac{u\, \sqrt{ s } }{2 }   \, \Bigl \{ \, 
R_- \left ( \gamma^2_x x^2 + \frac{ y^2}{\gamma_x^2}\, \right )
- 
R_+ \left ( \gamma^2_q q^2 + \frac{ p^2}{\gamma_q^2}\, \right ) \Bigr \}
$$
with
$$
R_- = \sqrt{ 2s + \sigma^2 -{\Delta}/{2}  }\, ,\quad R_+ = \sqrt{ 2s + \sigma^2 +{\Delta}/{2}  }\, ,
$$
(recall that $s = v +1/2$) and
\begin{equation}
\gamma_x^2 = \sqrt{ \frac{4 + \sigma^2  }{ 2s + \sigma^2 - {\Delta}/2 }} 
\, ,\,\,\,
\gamma_q^2 = \sqrt{ \frac{4 + \sigma^2  }{ 2s + \sigma^2 + {\Delta}/2 } }
\, .
\label{sigmaxq}
\end{equation}
The crucial condition that $2s + \sigma^2 -{\Delta}/{2} >0$ can be easily verified.
The eigenstates are defined as product states $|m,n\rangle = \Psi_m (x) \Psi_n (q)$
such that, for each oscillator,
\begin{eqnarray}
\Bigl ( \gamma^2_x x^2 + { y^2}/{\gamma_x^2}\, \Bigr ) \Psi_m (x) &=& (2m+1) \Psi_m (x) \, ,
\nonumber \\ 
\Bigl ( \gamma^2_q q^2 + { p^2}/{\gamma_q^2}\, \Bigr ) \Psi_n (q) & = &  (2n+1) \Psi_n (q)\, ,
\nonumber
\end{eqnarray}
$\Psi_m (x)$ and $\Psi_n (q)$ being standard harmonic-oscillator eigenfunctions.
The resulting spectrum reads
\begin{equation}
E(m,n)= \frac{u\, \sqrt{s} }{2 }  
\, \Bigl \{ \, R_-  \bigl ( \, 2\,m  + 1\, \bigr )
- 
R_+ \bigl ( 2\, n + 1 \, \bigr ) 
\Bigr \}
\label{E1}
\end{equation}
with $m, n \in {\mathbb N}_0$,
and the initial Hamiltonian satisfies the
eigenvalue equation $H_0 |E(m,n) \rangle = E(m,n) \, |E(m,n) \rangle$
where, in view of the preceding diagonalzation process, one has
\begin{equation}
\begin{cases}
& H_0 = S^+_\alpha U^+_\theta  D^+_\phi H_3 D_\phi U_\theta S_\alpha \, ,
\cr
& {\-} \cr
& 
|E(m,n) \rangle = S^+_\alpha U^+_\theta D^+_\phi |m,n \rangle \, .
\cr
\end{cases}
\label{AA1}
\end{equation}
%
%
%
\subsection{Diagonalization of case $v< -1/2$}
\label{3B}

\begin{figure}
\begin{centering}
\includegraphics[width= 7.0 cm]{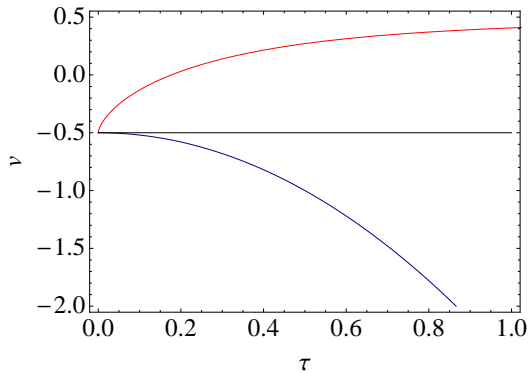}
\caption{
\label{fig1} 
(Color online)
The $\tau v$ plane features two regions ${\cal D}_+$ and ${\cal D}_-$ 
where the diagonalization process 
reduces the trimer Hamiltonian to a double-oscillator system with a discrete spectrum. 
${\cal D}_+$ is the upper domain
bounded from below by the red (upper) curve $\Gamma_+$ defined by $f(v)$ (see equation (\ref{area1})).
${\cal D}_-$ is delimited by the black straight line $\Gamma_0$ ($v=-0.5$)
and bounded from below by the blue (lower) curve $\Gamma_-$ defined by $g(v)$ 
(see equation (\ref{area2})).
}
\end{centering}
\end{figure}

\noindent
In this regime, involving a large central barrier, the diagonalization process can be performed 
by means of four subsequent steps. These correspond to the four unitary transformations 
forming $W = D_\phi S_\beta U_2 S_\alpha $ whose action allows one to obtain the 
new diagonal Hamiltonian $H_4' = W\, H_0 W^+$. 
The progressive action of such transformations is discussed in appendix \ref{app3}.
One obtains
$$
H_4' =
\frac{u\, \sqrt w}{2 }
\left [ 
{\cal R}_+ \left ( \nu_x^2 \, x^2  +\, \frac{ y^2}{\nu_x^2} \,   \right  )
-{\cal R}_-  \left ( \nu_q^2 \, q^2 \, + \, \frac{ p^2 }{\nu_q^2 } \, \right ) \right ]
$$
with
${\cal R}_\pm  =\sqrt{\eta^2 +2 w \, \pm D } $, $D = \, 2\sqrt { \eta^2 (w+1) + w^2 }$,
\begin{equation}
\eta = 2 \sqrt 2 \tau/\sqrt w\, , \quad w =|v|-1/2
\label{eta}
\end{equation}
and
\begin{equation}
\nu_x^2 = \frac{ \sqrt{ \eta^2-4 }}{\sqrt{\eta^2 +2 w \, + D } } 
\, ,\,\,\,
\nu_q^2 = \frac{ \sqrt{ \eta^2-4 }}{\sqrt{\eta^2 +2 w \, - D } }
\, .
\label{nuxq}
\end{equation}
Similar to the preceding case where $v >-1/2$,
the eigenstates are product states $|k,\ell \rangle = \Phi_k (x) \Phi_\ell (q)$
formed by harmonic-oscillator eigenfunctions such that
\begin{eqnarray}
\Bigl ( \nu^2_x x^2 + { y^2}/{\nu_x^2}\, \, \Bigr ) \Psi_k (x) &=& (2k+1) \Psi_k (x) \, ,
\nonumber \\ 
\Bigl ( \nu^2_q q^2 + { p^2}/{\nu_q^2}\, \Bigr ) \Psi_\ell (q) & = &  (2\ell+1) \Psi_\ell (q)\, ,
\nonumber
\end{eqnarray}
The resulting spectrum reads
\begin{equation}
E(k,\ell)= \frac{u\, \sqrt{w} }{2 }  
\, \Bigl \{ \, {\cal R}_+  \bigl ( \, 2\,k  + 1\, \bigr )
- 
{\cal R}_- \bigl ( 2\, \ell + 1 \, \bigr ) 
\Bigr \} \, .
\label{E2}
\end{equation}
where $k, \ell \in {\mathbb N}_0$. 
The original Hamiltonian $H_0$ satisfies the
eigenvalue equation $H_0 |E(k,\ell) \rangle = E(k,\ell) \, |E(k,\ell) \rangle$
where
\begin{equation}
\begin{cases}
& H_0 =S^+_\alpha U^+_2  S^+_\beta D^+_\phi H_4' D_\phi S_\beta U_2 S_\alpha \, ,
\cr
& {\-} \cr
& 
|E(k,\ell) \rangle = S^+_\alpha U^+_2  S^+_\beta D^+_\phi |k,\ell \rangle \, .
\cr
\end{cases}
\label{AA2}
\end{equation}
The range of validity of transformations $S_\alpha$ and $S_\beta$ in terms of
parameter $v$ defines the region ${\cal D}_-$ of parameter space $v \tau$ in which, for $v<-1/2$,
the diagonalization process succeeds. The actions of $S_\alpha$ 
is well defined if $v < -1/2$ (this ensures that $e^\alpha \in \mathbb R$),
while $S_\beta$, for a given $v$, requires that $\eta > 2$, namely,
\begin{equation}
\tau > g(v) \, , \quad g(v) \equiv 
\sqrt{ |v| -1/2}/{\sqrt 2} \, ,
\label{area2}
\end{equation}
(see appendix \ref{app3}). We will denote curve $\tau = g(v)$ and 
the straight line $v = -1/2$ with $\Gamma_-$ and $\Gamma_0$, respectively.
Thus domain ${\cal D}_-$ corresponds to the region bounded by 
$\Gamma_-$ from below and by $\Gamma_0$ from above (see the caption of Fig. \ref{fig1}). 
This result confirms that the regions of plane $\tau v$ where the
current diagonalization scheme is effective correspond to the regions where classical
oscillations are stable.

%

\section{ structure of the energy spectrum and classical instability}
\label{quattro}

\subsection{Spectrum of case $v> -1/2$}
\label{4A}

Domain ${\cal D}_+$ described by inequalities (\ref{area1})
reproduces in Fig. \ref{fig1} the first of the two regions 
of the stability diagram relevant to the CDW regime \cite{jpa42} 
in which classical trajectories with initial condition close to the CDW
solution are dynamically stable. The property that CDW classical states are stable 
thus corresponds, quantum-mechanically, to the fact that the diagonalization process reduces the 
system to a simple double oscillator.

The dependence of energy spectrum (\ref{E1}) on parameters $\tau$ and $v$ allows one to identify
two regimes characterized by the inequalities
$$
\tau << \, 1\, ,\qquad \Delta << 1, 
$$
respectively, in which the spectrum manifests significant changes.
Region ${\cal D}_+$ of Fig. 1 shows that,
for a given value of $v$ in the interval $[-1/2,\, 1/2]$, parameter $\tau$ ranges in $[0, f(v)]$,
where $\tau = f(v)$ defines boundary $\Gamma_+$ of ${\cal D}_+$.
While the first regime 
corresponds to values of $\tau$ close to the vertical axis $v $ (interwell tunneling inhibited),
the second regime, where $\Delta \to 0$, corresponds to approaching 
boundary $\Gamma_+$ from the left, namely, $\tau \to (f(v))^-$. Thus $\Delta \to 0$
drives the approach to the region where the classical instability occurs.

{\it Regime $\tau <<1$}. 
In the weak-tunneling case, by exploiting the Taylor expansion of $R_\pm$ to the second order
in $\tau$, one obtains the expression
$$
E(m,n)= \frac{u\sqrt s}{2 }  
\left [ \frac{ 2 {\sqrt 2}\, \tau  }{ s } \bigl ( 2m  + 1 \bigr )
- R_+  \bigl ( 2n +1 \bigr ) \right ]\, ,
$$
with 
$$
s= \sqrt{ v +1/2 }  \, ,\,\,\, R_+ \simeq \sqrt{ 4s\, + \frac{ 16 v \, \tau^2 }{s^2 } }\, ,
$$
showing that quantum number $n$ describes the large-scale energy changes while quantum number $m$
describes the {\it fine structure} of the spectrum. The reference case, of course, is represented by 
$E(0,0)$, the energy of the quantum states $|0, 0\rangle$ corresponding to the minimum deviation from 
the pure CDW configuration.

The presence of the parabola, represented by $v >0$, diminishes the fine-structure level separation
with respect to the case $v=0$ involving no confinement. 
%
A more apparent effect occurs for $v < 0$ where the attractive potential of the 
central well becomes a potential barrier.
In this case the fine-structure level separation obtained with small $\tau$ is contrasted by 
factor $1/\sqrt {s}$. Decreasing $v$ by maintaining $\tau$ constant shows that such an effect
is maximum when one approaches boundary $\Gamma_+$ from above.

{\it Regime $\Delta <<1$}. 
The two oscillators tend to become identical ($\gamma_x- \gamma_q \to 0$) and the
relevant energy levels are almost indistinguishable. This determines a {\it macroscopic change}
of the energy spectrum of $H_0$ which becomes visible by effecting, with $|v| < 1/2$, the Taylor 
expansion of $R_\pm$ in terms of variable $\Delta$. The resulting spectrum has the form
$$
E(m,n)= \, \frac{u\, \sqrt{ s} }{2 }  
\, \left [ \, 2R \, ( m - n )
-\frac{\Delta\, ( m+n+1) }{2 R  }
\right ]
$$
with $R := \sqrt{ 2v +1 + \sigma^2  } = R_\pm$ for $\Delta =0$. 
%
Eigenvalues $E(m,n)$ feature a band structure described by $h = m-n$ and a {\it fine structure} 
controlled by parameter $\Delta$ and the composite quantum number $m+n+1$. 
The choice $m=n$ describes the band exhibiting the minimal deviation
from the pure CDW configuration.

For $\Delta \to 0$ the fine structure of each band becomes infinitely dense since the interlevel
distance tends to zero. In this regime the behaviors of the two oscillators are 
{\it strongly correlated}
in that small energy changes requires the simultaneous change of $m$ and $n$ in order
to preserve $h$.
All the effects described so far disappear at $v=1/2$ (and, more in general, for $v > 1/2$) 
since parameter $\Delta $ no longer vanishes.
Also, the range of $\tau$ becomes unlimited consistent with the fact that 
$\tau$ does not cross the unstable-regime boundary $\Gamma_+$.

\subsection{Spectrum of case $v< -1/2$}
\label{4B}

The present regime $v < -1/2$ features as well two significant limiting cases
occurring in the proximity of boundaries $\Gamma_-$ and $\Gamma_0$. These are
$$
\eta -2 <<1 \, , \qquad |v| -1/2 << 1  
$$
(see definition (\ref{eta})) that correspond
to approaching 1) the curve $g(v)$ from the right, at a given $v$,
and 2) the straight line $v= -1/2$ from below, at a given $\tau$, respectively.
Hence,
the approach to the regions where the classical instability crops up
is driven by either $\eta -2 \to 0$ or $|v| \to 1/2$.

{\it Regime $\eta -2 <<1$}. One easily checks that $\eta \to 2^+$ amounts to effecting 
the limit $\tau \to g(v) \equiv \sqrt{w/2}$.
By substituting $\tau = \eta \sqrt{w}/ (2\sqrt 2)$ in ${\cal R}_\pm$, and considering
the Taylor expansion in $\eta -2$, one obtains
$$
{\cal R}_+ \simeq 2 \sqrt {2+w}\, ,\quad {\cal R}_- \simeq \, 2 \sqrt{ \eta -2} \, \rho   
$$
with $\rho = \sqrt{(2w+3)/(w+2)}$ entailing that
$$
E(k,\ell)=  \, u \sqrt{w}   
\, \Bigl [  
\sqrt {2+w}\, \bigl ( \, 2\,k  + 1\, \bigr )
- 
\sqrt{ \eta -2} \,\rho \bigl ( 2\, \ell + 1 \, \bigr ) 
\Bigr ] \, .
$$
Quantum number $k$ represents the band index,
while the fine structure of the spectrum is controlled by number $\ell$. For 
$\eta \to 2^+$, similar to the case $\tau \to 0$ of subsection \ref{4A}, 
the spectrum acquires an almost continuous character.

{\it Regime $|v| -1/2 <<1$}. In this case, effecting the limit $v \to (-1/2)^-$ 
in equation (\ref{E2}) implies that
$$
E(k,\ell)= \, u  
\, \Bigl [ \,  2{\sqrt 2} \tau \bigl ( \, k -\ell \, \bigr )
+
{\sqrt w } \bigl ( \, k  + \ell + 1\, \bigr ) \Bigr ] \, ,
$$
where the approximation to the first order in $w$
$$
\sqrt{w}  {\cal R}_\pm \simeq 
2{\sqrt 2} \tau \left ( 1 \pm \frac{\sqrt w}{2{\sqrt 2} \tau} \right )
= 2{\sqrt 2} \tau \pm \sqrt w
$$
has been used. Spectrum (\ref{E2}) thus undergoes a {\it macroscopic change}
with energy levels forming
a band structure described by index $h = k-\ell$,
and a $w$-dependent fine structure described by index $k + \ell$. 
The separation between two subsequent levels tends to zero for $w \to 0$. 
Transitions in which $k\to k+r$, $\ell \to \ell +r$ are thus favoured in that 
small energy changes are involved. Similar to the case $\Delta <<1$ of section \ref{4A},
the two oscillators appear to be {\it strongly correlated} in that $\Delta k \equiv \Delta \ell$.

\subsection{Transition to a continuous spectrum}
\label{4C}

Transformations $W= D_\phi U_\theta S_\alpha$ and $W= D_\phi S_\beta U_2 S_\alpha$
(see formulas (\ref{AA1}) and (\ref{AA2}), respectively) no longer work
at the boundaries $\Gamma_+$, $\Gamma_0$ and $\Gamma_-$ of the relevant stability domains.
In particular, in the case $v> -1/2$, the action of $D_\phi$ in $W$ is defined
for arbitrarily large values of $\phi$ as shown by equation (\ref{th1}). The latter shows that
${\rm th} (\phi) \to 1$ for $\tau \to f(v)$ meaning that $\Gamma_+$ is approached 
(but not reached) from below.
Likewise, for $v <-1/2$, the range of parameters $\alpha$ and $\beta$
(see appendix \ref{app3}) relevant to $S_\alpha$ and $S_\beta$, respectively, allows one
to get closer and closer to boundaries $\Gamma_0$ and $\Gamma_-$ without reaching them.

At such boundaries the spectrum displays a structural change corresponding to the
fact that the diagonalization of $H_0$ yields an operator pertaining to a sector of
the dynamical algebra disjoint from
the one where $H_0$ reduces to a simple two-oscillator model. 
In the new sector the discrete character of the spectrum is lost. A paradigmatic
example of such an effect is supplied by the harmonic oscillator with time-dependent
parameters \cite{AP96}.

Hamiltonian $H_0$ at the boundary $\Gamma_0$ where $v \equiv -1/2 $ well exemplifies this
situation. In this case, $H_0$ reduces to
$H_0 = u [ x^2 - 2{\sqrt 2} \tau (xq+yp) ]$.
The action of transformation ${ M} = e^{-i\varphi xp }$ on $H_0$ giving $H_0' = { M} H_0 { M}^+$
followed by the unitary transformation $x \to y$, $y \to -x$
takes $H_0$ into the form
$$
H_0''
= u \left  [ \frac{1}{2} ( y^2 +  p^2 ) - g (xp -yq) \right ]
$$
which exhibits the two-dimensional Laplacian $y^2 +  p^2$ together with 
the generator  $L_3 = xp -yq$ of planar rotations. After noting that such operators commute
with each other, one easily observes that, despite $L_3$ has a discrete spectrum, 
the energy spectrum features a continuous character
due to the presence of Laplacian. 
Since the same result can be shown to occur at boundaries $\Gamma_-$
and $\Gamma_0$, then we conclude that at the border separating
(classically) stable from unstable regimes the energy spectrum acquires a continuous
character. 

The correct interpretation of such an effect is twice.
At the boundaries $\Gamma_+$, $\Gamma_-$ and $\Gamma_0$ of ${\cal D}_\pm$
the approximation inherent in model (\ref{H0}), involving microscopic modes $A$ and $C$,
is no longer valid. This is obvious noting that the 
spectrum of quantum trimer is discrete by definition. The onset of a continuous spectrum
is thus an artifact of the momentum-mode picture when this is used outside its domains
of validity ${\cal D}_\pm$.
On the other hand, the emergence of a continuous spectrum just amplifies the effect
observed in the proximity of $\Gamma_+$, $\Gamma_-$ and $\Gamma_0$ which consists in 
the vanishing of energy-level separation (see subsections \ref{4A} and \ref{4B}).
In this sense, the continuous spectrum is a dramatic manifestation of the
transition to the classically-unstable regions.

%

\section{Delocalization effects}
\label{cinque}
 
The discussion of sections \ref{4A} and \ref{4B} shows that
both regimes $v> -1/2$ and $v< -1/2$ feature limiting cases
where a suitable choice of parameters $v$ and $\tau$ allows one to
approach the boundaries of the domains classically involving stable
oscillations. 
The significant change of the spectrum structure observed in
such cases is accompanied by a dramatic change of the localization properties
of the system. 
To see this one must consider deviations
$\Delta^2_f = \langle f^2 (t) \rangle - \langle f (t) \rangle^2$ where $f$ represents one 
of canonical  operators $x, y, q$, and $p$, and
$$
\langle f^r (t)\rangle = \langle E| f^r (t) |E\rangle
= \langle E| R^+ f^r R  |E\rangle\, ,
$$
with $r= \, 1,2$. The action of propagator $R = e^{-itH_0/\hbar}$ on energy eigenstates $|E\rangle$
allows one to considerably simplify this calculation.
Formulas (\ref{AA1}) and (\ref{AA2}) show that $H_0 = W^+ H_d W$ and 
$|E (\Omega) \rangle = W^+ |\Omega \rangle$
in which
$$
W= D_\phi U_\theta S_\alpha \, , \quad  W= D_\phi S_\beta U_2 S_\alpha \, ,
$$
In the two cases $v>-1/2$ and $v<-1/2$, one has
$H_d = H_3$, $H_d = H_4'$, and $|\Omega \rangle = |m,n\rangle$, $|\Omega \rangle = |k,\ell\rangle$, 
respectively. As a consequence the calculation of expectation values
$\langle f^r (t)\rangle $ reduces to
$$
\langle \Omega | W\, R^+ f^r R \, W^+ |\Omega \rangle 
= \langle \Omega | W\, f^r W^+ |\Omega \rangle\, ,
$$ 
being $H_0 =  W^+ H_d W$ in propagator $R(t)$. One easily shows that 
$\langle \Omega |\, W f W^+  |\Omega \rangle = 0$ 
so that the determination of $\Delta^2_f$ amounts to calculating
$$
\Delta^2_f = \langle n,m | D_\phi U_\theta S_\alpha \, f^2 
S^+_\alpha U^+_\theta D^+_\phi |m,n \rangle \, ,
$$
$$
\Delta^2_f = \langle \ell,k | D_\phi S_\beta U_2 S_\alpha \, f^2 
S^+_\alpha U^+_2  S^+_\beta D^+_\phi  |k,\ell \rangle \, ,
$$
with $f= x,y, q, p$. The derivation of such formulas is discussed in Appendix \ref{app4}. 

Let us consider first
the approach to boundary $\Gamma_+$ characterized by $\Delta \simeq 0$. In the proximity
of $\Gamma_+$ one finds that
$\tau^2 \simeq f^2(v) = s^3/[8(1-s)]$ and $\sigma^2 \simeq s^2/(1-s)$ while 
$R_\pm \simeq R = (2s + \sigma^2)^{1/2}$ and $F_\pm \simeq F = s(2-s)$. 
Hence, equations (\ref{delxy1}) and (\ref{delqp1}) reduce to
$$
\Delta^2_x \simeq \frac{2R^3 \sqrt s}{ (4+\sigma^2) \Delta} (m+n+1)\, ,\,\,\,
\Delta^2_y \simeq \frac{2F }{ R {\sqrt s} \Delta} (m+n+1)\, ,
$$
$$
\Delta^2_q \simeq \frac{2RF \, (m+n+1) }{ (4+\sigma^2) {\sqrt s} \Delta} \, ,\,\,\,
\Delta^2_p \simeq \frac{2R {\sqrt s}  }{ \Delta} (m+n+1)\, ,
$$
clearly showing how such squared deviations diverge owing to the factor $1/\Delta$. For
$s \ge 1$ ($\leftrightarrow v\ge 1/2$) parameter $\Delta$ does not vanish any longer 
whatever value is assumed by $\tau$. The delocalization effect is thus restricted to
neighborhood of $\Gamma_+$.

Concerning the boundaries of ${\cal D}_-$, the approach to $\Gamma_-$ 
(blue curve in Fig. 1) discloses a somewhat different diverging behavior
of squared deviations $\Delta_f^2$. 
In particular, based on formulas (\ref{delxy2}), (\ref{delqp2}) and (\ref{lim2}), one gets
$$
\Delta^2_x \simeq \frac{(2\ell+1) \sqrt w}{ 2 \sqrt{\epsilon} \sqrt{w+2 } } \, ,\,\,\,
\Delta^2_y \simeq  \frac{(2k+1)}{ 4 \sqrt{w (w+2) } } \, ,
$$
$$
\Delta^2_q \simeq \frac{(2\ell+1) }{ 2 \sqrt{\epsilon } \sqrt{w (w+2)} } \, ,\,\,\,
\Delta^2_p \simeq \frac{(2k+1) \sqrt w }{ 4 \sqrt{w+2} } \, ,
$$
in the limit $\epsilon = \eta^2-4 \to 0$, where factor $1/\sqrt \epsilon$ is responsible
for the delocalization effect. The latter affects only position operators $x$ and $q$
without involving $y$ and $p$.

Finally, the limit $w \to 0$, describing the approach to
the straight line $\Gamma_0$ defined by $v = -1/2$, supplies a third interesting situation.
In this case
$$
\Delta^2_x = \Delta^2_p \simeq \frac{\sqrt w}{ 2 }(k + \ell+1) 
\, ,\,\,\,
\Delta^2_y = \Delta^2_q \simeq  \frac{k + \ell+1}{2 \sqrt{w} } \, ,
$$
involve an evident squeezing effect where the delocalization issuing from 
$\Delta^2_y$ and $\Delta^2_q$ is compensated by the localization effect 
concerning $\Delta^2_x$ and $\Delta^2_p$, respectively.


\section{Conclusions}
\label{sei}

We have studied the quantum-mechanical properties of the Bose-Hubbard trimer 
in the CDW regime.
The interesting feature of this regime is that it exhibits both stable
and unstable (classical) regimes whose dependence on interaction parameters is
completely known. This has given the possibility to study the relation
between the stability character of trimer and its quantum properties.

In Section~\ref{due}, we have introduced an alternative description in terms 
of momentum-like modes which allows one to identify the classical (microscopic) 
variables involved by the quantization process. In Section~\ref{tre} we have enacted 
the procedure for diagonalizing the trimer Hamiltonian based on the knowledge of the 
model dynamical algebra. Within the momentum-mode picture, 
this is found be the algebra sp(4).

The implementation of this procedure reduces the model to a double-oscillator system
and allows us to identify the regions ${\cal D}_+$ and ${\cal D}_-$ where the spectrum
is discrete. The nice analytic result emerging from this part is that the derivation
of (the boundaries of) ${\cal D}_+$ and ${\cal D}_-$ exactly reproduces the 
regions where the trimer is classically stable in the CDW regime. 

The discussion of section~\ref{quattro} makes evident the significant changes of
spectrum structure when approaching the boundary of stability domains 
${\cal D}_+ $ and ${\cal D}_-$. 
Various delocalization effects, discussed in section~\ref{cinque}, confirm
the considerable sensitivity of the quantum properties of the system from the proximity
to the boundaries of ${\cal D}_+ $ and ${\cal D}_-$.
All such effects can be interpreted as the hallmark of the transition 
to the unstable regions of classical CDW dynamics whose most dramatic manifestation
is the onset of a continuous spectrum 
when the domain of validity of the momentum-mode picture is left.

Future work will be focused on extending the investigation of the quantum properties of 
trimer in more complex regimes (such as the dimeric and nondimeric regimes 
studied in Refs. \cite{BFPprl90} and \cite{jpa42}) where the transition to unstable 
oscillations can be controlled through the model parameters. 
We hope to include various macroscopic dynamical effects reflecting parameter-dependent 
structural changes of the phase space
so far examined only for the classical trimer \cite{jpa42}.
\acknowledgments

The author wishes to thank P. Buonsante and R. Franzosi for stimulating discussions.
This work has been partially supported by the M.I.U.R. project {\it Collective quantum 
phenomena: From strongly correlated systems to quantum simulators} (PRIN 2010LLKJBX). 
%
%

\appendix
\section{Collective-mode quantum picture}
\label{app1}
Similar to classical modes (\ref{ABC}) the new quantume modes are defined by
$A = (a_1+a_3)/ \sqrt 2$, $B = (a_1 -a_3)/ \sqrt 2$ and $C= a_2$.
By substituting the latter in Hamiltonian (\ref{qHam}) and observing that
$$
(a_1^+ a_1)^2 + (a_3^+ a_3)^2=
\frac{1}{2} (n_A +n_B)^2 + \frac{1}{2} (A^+B + B^+A)^2
$$
with $n_A = A^+A$ and $n_B = B^+B$, one finds
$$ 
H= \frac{U}{4} \Bigl [ n^2_A + n^2_B + 4n_A n_B   + (A^+)^2 B^2 + (B^+)^2 A^2  
$$
$$ 
+2 n^2_C + n_A +n_B \Bigr ] - \frac{U}{2} N -V n_C - {\sqrt 2} T ( A^+ C + C^+ A) . 
$$
This Hamiltonian must be transformed in order to get the quantum version of ${\cal H}_f$. 
Quantum-mechanically, 
the time-dependent canonical transformation (\ref{abc}) corresponds the unitary transformation
${\cal U}_t = \exp [ i \varphi(t) N ]$ with $\varphi(t) = \varphi- ut/\hbar$ and $N = n_A +n_B +n_C$ 
whose action takes Hamiltonian $H = H (A,B,C)$ into Hamiltonian $H_f = H -u N$. This result
is achieved by changing the Hilbert-space basis through the action of ${\cal U}_t$, 
namely, by setting
$| \Psi \rangle  = {\cal U}_t | \Phi \rangle $ in the Schr\"ondinger problem
$i \hbar \partial_t | \Psi \rangle  = \, H | \Psi \rangle$ based on the $ABC$ picture.
The latter becomes
$i \hbar  (\partial_t {\cal U}_t ) | \Phi \rangle
+i \hbar  {\cal U}_t \partial_t | \Phi \rangle  = \, H \, {\cal U}_t | \Phi \rangle$
giving
$$
i \hbar  \partial_t | \Phi \rangle  = \, \Bigl [ 
-i \hbar   {\cal U}_t^+ (\partial_t {\cal U}_t ) + {\cal U}_t^+ H\, {\cal U}_t \Bigr ]\, | \Phi \rangle \, .
$$
Then
$$
H_f = -i \hbar   {\cal U}_t^+ (\partial_t {\cal U}_t ) + {\cal U}_t^+ H  {\cal U}_t
= H + \hbar N {\dot \phi} = H - u\, N \, ,
$$
being $[N, H]=0$. The invariance of Hamiltonian $H$, entailing that ${\cal U}_t^+ H(A,B,C)\, {\cal U}_t = H(A,B,C)$,
directly follows from the transformation formulas
\begin{equation}
{\cal U}_t X \,{\cal U}^+_t = X  e^{-i\varphi (t)} \, ,\quad X= A,B,C\, .
\label{abcQ}
\end{equation}
The action described by (\ref{abcQ}) is equivalent to introducing $a$, $b$ and $c$ within the classical
description. Conversely, the latter are not required in the quantum picture since transformations
(\ref{abcQ}) introduce in a direct way the time-dependent exponential factor $e^{-i\varphi (t)}$. 
In the present scheme, for example, the expectation values 
$\langle \Psi |A| \Psi \rangle$ (relevant to the dynamics governed by $H$) and 
$\langle \Phi |A| \Phi \rangle$ (relevant to the dynamics governed by $H_f$)
correspond to the classical variables $A$ and $a$, respectively, since
$$
\langle \Psi |A| \Psi \rangle = \langle \Phi |{\cal U}^+_t A \, {\cal U}_t | \Phi \rangle
= e^{i\varphi} \langle \Phi | A | \Phi \rangle \, .
$$
The last step consists in replacing $B$
with ${\sqrt N} + \xi$. The substitution of the latter in $n_B$ and $n_B^2$ 
gives $n_B= N + {\sqrt N} (\xi +\xi^+) + \xi^+ \xi $ and
\begin{eqnarray}
n^2_B &=&  (N + {\sqrt N} (\xi +\xi^+) + \xi^+ \xi )^2
\\
& \simeq & N^2 + N(\xi +\xi^+)^2 + 2 N {\sqrt N} (\xi +\xi^+) + 2N \xi^+ \xi \, ,
\label{n2}
\nonumber
\end{eqnarray}
where terms such as $(\xi^+ \xi)^2$, $\xi^2 \xi^+$ and $(\xi^+)^2 \xi$ have been
neglected. This omission is justified by the fact that mode $\xi$, as well as $A$ and $C$, 
represent weakly populated modes. Hence the expectation values of $(\xi^+ \xi)^2$ and $\xi^2 \xi^+$
represent pertubative terms that can be neglected. For the same reason
terms $(A^+)^2 B$ and  $A^2 B^+$ can be eliminated as well as terms $n_A^2$ and $n_C^2$. 
Then $H_f$ reduces to
\begin{eqnarray}
H_f & \simeq & \frac{U}{4} \Bigl [ n^2_B + 4n_A n_B + B^2 (A^+)^2 + A^2 (B^+)^2 \Bigr ]
\nonumber 
\\
&-& u (n_A + n_B + n_C)  -V n_C - {\sqrt 2} T ( A^+ C + C^+ A)\, ,
\nonumber
\end{eqnarray}
%
%
since $U (n_A + n_B  + n_C )/2$ and $U (n_A + n_B )/4$ are negligible
with respect to $u (n_A + n_B  + n_C )$, being $u = UN/2 >> U$.
Implementing the substitution $B= {\sqrt N} + \xi$ gives
\begin{eqnarray}
H_f & \simeq & - \frac{u}{2}N + u (\xi +\xi^+)^2 + u n_A + \frac{u}{2} \Bigl [ (A^+)^2 + A^2 \Bigr ]
\nonumber 
\\
&-& (V+u) n_C - {\sqrt 2} T ( A^+ C + C^+ A)\, .
\nonumber
\end{eqnarray}
%


\section{Canonical form of algebra sp(4)}
\label{app2}

By expressing bosonic operators in terms of canonical variables $A = ({x+iy})/{\sqrt 2}$ and 
$C =( {q+ip})/{\sqrt 2}$ the canonical version of algebra sp(4) is given by
\begin{equation}
\begin{cases}
&  
J_1 = \frac{A^+ C + C^+ A}{2} = \frac{xq+yp}{2}
\cr
& {\-} \cr
&
J_2 = \frac{A^+ C - C^+ A}{2i} = \frac{xp -qy}{2}
\cr
& {\-} \cr
& 
J_3 = \frac{A^+ A - C^+ C}{2} = \frac{x^2 +y^2 - q^2-p^2 }{4}
\cr
& {\-} \cr
& 
J_0 = \frac{A^+ A + C^+ C +1}{2} = \frac{x^2 +y^2 + q^2 +p^2 }{4}
\cr
\end{cases}
\, .
\label{JJJ}
\end{equation}
\begin{equation}
\begin{cases}
&  
K_3 = \frac{ C A + A^+ C^+}{2} = \frac{xq - yp}{2}
\cr
& {\-} \cr
&
K_2 = \frac{A^{2} - A^{+2} + C^{2} - C^{+2} }{2i} = \frac{xy+yx + qp+ pq}{4}
\cr
& {\-} \cr
& 
K_1 = \frac{ C^{2} + C^{+2} - A^{2} - A^{+2} }{2} = -\frac{x^2 - y^2 - (q^2 - p^2) }{4}
\cr
\end{cases}
\, .
\label{QQQ}
\end{equation}
\begin{equation}
\begin{cases}
&  
Q_1 = \frac{A^{+2} -A^{2} + C^{2} - C^{+2}}{2i} = - \frac{xy- qp }{2}
\cr
& {\-} \cr
&
Q_2 = \frac{ A^{2} +A^{+2}  + C^{2} + C^{+2}  }{2} = -\frac{x^2 - y^2 + q^2 - p^2 }{4}
\cr
& {\-} \cr
& 
Q_3 = \frac{A C - A^+ C^+ }{2i} = \frac{xp + qy }{2}
\cr
\end{cases}
\, .
\label{KKK}
\end{equation}
The unitary transformations whereby canonical operator $x$, $y$, $q$ and $p$
can be  modified
and thus $H_0$ can be diagonalized are essentially three. The latter are defined by
equations (\ref{Ualfa}), (\ref{Utheta}) and (\ref{Uphi}) representing the action 
of $S_\alpha$ (squeezing transformation), $U_\theta$ (rotation) and 
$D_\phi$ (hyperbolic transformation), respectively.
%

\section{Diagonalization scheme for $v< -1/2$}
\label{app3}
The first unitary transformation is again a squeezing transformation 
(\ref{Ualfa}) $S_\alpha = \exp (-i\alpha Q_1)$ which modifies $H_0$ as follows 
$$
H_1'= 
S_\alpha H_0 \, S_\alpha^+ =
u {\sqrt{w}}  \Bigl [ \, x^2 + q^2 + w \,p^2 
- \eta \, (xq\, + yp ) \, \Bigr ]\, ,
$$
with $w = |v|-1/2$ and $ \eta \equiv 2\tau (2/w)^{1/2} $,
provided the condition $e^\alpha = \sqrt w$ is satisfied.
The second transformation is a simple rotation $U_2 = \exp (-2i\theta J_2)$
with $\theta = \pi/4$ (see equation  (\ref{Utheta})) entailing
$$
\begin{cases}
& U_2 x U^+_2 = \frac{x + q}{\sqrt 2} , \,\, U_2 y U^+_2 = \frac{y  + p}{\sqrt 2} ,
\cr
& {\-} \cr
& U_2 q U^+_2 =  \frac{q -x}{\sqrt 2}, \,\, U_2 p U^+_2 =  \frac{p-y}{\sqrt 2},
\cr
\end{cases}
$$
and
\begin{eqnarray}
H_2'  & = & U_\theta H_1' \, U_\theta^+ =
\frac{u}{2} {\sqrt{w}}  \Bigl [ \, ( \eta+2) x^2 - ( \eta-2) q^2 
\nonumber 
\\
&+&  (w-\eta) \,p^2 +\, (w+\eta) \, y^2 -2wyp \, \Bigr ] \, .
\nonumber
\end{eqnarray}
To obtain terms $x^2$ and $q^2$ with the same coefficients, we implement a second 
squeezing transformation $S_\beta = \exp (+i\beta Q_1)$ leading to
\begin{eqnarray}
&H_3' &  = S_\beta H_2' \, S_\beta^+ =
\frac{u}{2} {\sqrt{w}}  \Bigl [ \, \sqrt{\eta^2-4} \, ( x^2 -q^2)  
\nonumber 
\\
&+& \frac{\eta^2 +2 w}{\sqrt{ \eta^2-4 }}  (y^2 - p^2) 
+ \frac{\eta (w+2)}{\sqrt{ \eta^2-4 }}  (y^2 + p^2)- 2wyp \, \Bigr ]
\, , 
\nonumber
\end{eqnarray}
where parameter $\beta$ must fulfil the condition
$$
e^{2 \beta} =\, {\sqrt{(\eta+2)/(\eta -2)}}\, . 
$$
The form of Hamiltonian $H_3'$ is suitable for applying hyperbolic-like transformations (\ref{Uphi})
that leave terms $x^2 -q^2$ and $y^2 -p^2$ unchanged. One finds
\begin{eqnarray}
&H_4' & =  S_\phi H_3' S_\phi^+  =
\frac{u}{2} {\sqrt{w}}  \Bigl [ \, \sqrt{\eta^2-4} \, ( x^2 -q^2) 
\nonumber 
\\
&+& 
\frac{\eta^2 +2 w}{\sqrt{ \eta^2-4 }} (y^2 - p^2) 
+
\frac{ 2\sqrt { \eta^2 (w+1) + w^2 } }{ \sqrt{ \eta^2-4 }}  ( y^2 + p^2 ) \Bigr ], 
\nonumber
\end{eqnarray}
where the condition
$$
\eta (w+2) \, {\rm sh} (2 \phi) \, 
+\,  w\, {\rm ch} (2 \phi) {\sqrt{ \eta^2-4 }} =0 
$$
--used to eliminate the mixed term $yp$--
provides ${\rm th} (2 \phi) \, = - \, {w \, \sqrt{ \eta^2-4 }}/{ [\eta (w+2)] }$
thereby determining parameter $\phi$. We thus obtain $H_4'$ in the form
$$
H_4' =
\frac{u\, \sqrt w}{2 }
\left [ 
{\cal R}_+ \left ( \nu_x^2 \, x^2  +\, \frac{ y^2}{\nu_x^2} \,   \right  )
-{\cal R}_-  \left ( \nu_q^2 \, q^2 \, + \, \frac{ p^2 }{\nu_q^2 } \, \right ) \right ]
$$
with ${\cal R}_\pm  =\sqrt{\eta^2 +2 w \, \pm D }$, $D = \, 2\sqrt { \eta^2 (w+1) + w^2 }$.
The explicit form of harmonic-oscillator frequencies $\nu_x$ and $\nu_q$ are defined by 
equations (\ref{nuxq}).
%
%

\section{Action of unitary transformation $W$ and calculation
of standard deviations}
\label{app4}

The combined action of (\ref{Ualfa}), (\ref{Utheta}) and (\ref{Uphi}) gives
$$
\begin{cases}
&  
\!\!\!\!  W x W^+ = e^\alpha  [ 
(C_\theta c_\phi + S_\theta s_\phi ) x +( C_\theta s_\phi+  S_\theta c_\phi ) q  ] 
\cr
& {\-} \cr
& 
\!\! \!\! W y W^+ = e^{-\alpha}  [
(C_\theta c_\phi - S_\theta s_\phi ) y +( S_\theta c_\phi -C_\theta s_\phi) p  ]
\cr
\end{cases}
,
$$
$$
\begin{cases}
&  
\!\!\!\!  
W q W^+ = e^{-\alpha} [
(C_\theta c_\phi - S_\theta s_\phi ) q +( C_\theta s_\phi -  S_\theta c_\phi ) x ]
\cr
& {\-} \cr
& 
\!\!\!\! 
W p W^+ = e^{\alpha} [
(C_\theta c_\phi + S_\theta s_\phi ) p - ( S_\theta c_\phi  +C_\theta s_\phi) y ]
\cr
\end{cases}
.
$$
where $W = D_\phi U_\theta S_\alpha$. This results clearly shows that 
$\langle n,m| W f W^+| m,n\rangle = 0$ when $f = x,y,q, p$.
By exploiting such transformation one easily calculates
deviations $\Delta^2_f = \langle n,m| W f^2W^+| m,n\rangle$
$$
\begin{cases}
&  
\!\!\!\!  
\Delta^2_x = \frac{1}{4} e^{2\alpha} 
\left ( (2m+1)\gamma_x^{-2} K_{++} +(2n+1) \gamma_q^{-2} K_{- +} \right )
\cr
& {\-} \cr
& 
\!\! \!\! 
\Delta^2_y = \frac{1}{4} e^{-2\alpha} 
\left ( (2m+1) \gamma_x^2 \, K_{+-}+ (2n+1) \gamma_q^2 \, K_{- -} \right )
\cr
\end{cases}
$$
$$
\begin{cases}
&  
\!\!\!\!  
\Delta^2_q = \frac{1}{4} e^{-2\alpha} 
\left ( (2n+1) \gamma_q^{-2} K_{+-} + (2m+1) \gamma_x^{-2} K_{--} \right )
\cr
& {\-} \cr
& 
\!\! \!\! 
\Delta^2_p = \frac{1}{4} e^{2\alpha} 
\left ( (2n+1) \gamma_q^2 \, K_{++} + (2m+1) \gamma_x^2 \, K_{- +} \right )
\cr
\end{cases}
$$
where
$ 
K_{\delta \mu}  = {\rm ch} (2\phi) + \delta \cos (2\theta) + \mu {\rm sh} (2\phi) \sin (2\theta)
$
with $\delta = \pm $, $\mu = \pm$, $e^{2\alpha} =\sqrt {v+1/2}$ and
%
$$
{\rm ch}  (2\phi)= { 2s\sqrt {4+\sigma^2} }/{\Delta}
\, , \,\,\,
{\rm sh} (2\phi)= {(3-2v)\sigma}/{ \Delta}\, ,
$$
$$
\cos (2\theta)= {2}/{\sqrt {4+\sigma^2}}
\, , \,\,\, 
\sin (2\theta)= {\sigma}/{\sqrt {4+\sigma^2}}
\, .
$$
One should recall that
$\sigma = 2\sqrt 2 \tau /\sqrt s$, $s=v+1/2$
and
$\Delta = 4\sqrt{s^3 +8\tau^2 (s-1) }/\sqrt{s}$, 
while 
$$
\gamma^2_x = {\sqrt {4+\sigma^2} }/{R_-}
\, ,\,\,
\gamma^2_q = {\sqrt {4+\sigma^2} }/{R_+}\, ,
$$
where $R_\pm = \sqrt {2s+ \sigma^2 \pm \Delta/2}$. In view of such 
definitions the final form of squared deviations $\Delta^2_f$
is given by
\begin{equation}
\begin{cases}
&  
\!\!\!\!  
\Delta^2_x = \frac{\sqrt s}{\Delta} \, 
\left [ (2m+1) \frac{ R^2_+ R_- }{4+ \sigma^2 } +(2n+1) \frac{ R^2_- R_+ }{4+ \sigma^2 } \right ]
\cr
& {\-} \cr
& 
\!\! \!\! 
\Delta^2_y = \frac{1}{\sqrt s} \, \left (
\frac{2m+1}{ R_-  \Delta } F_+
+
\frac{2n+1}{ R_+  \Delta } F_-  \right )
\cr
\end{cases}
\label{delxy1}
\end{equation}
with $F_\pm  = 2s+\sigma^2 (s-1) \pm \Delta/2$, and
\begin{equation}
\begin{cases}
&  
\!\!\!\!  
\Delta^2_q =
\frac{ 2n+1 }{{\sqrt s} (4+ \sigma^2) \Delta } R_+  F_+
+
\frac{ 2m+1 }{ {\sqrt s} (4+ \sigma^2) \Delta } R_-  F_- 
\cr
& {\-} \cr
& 
\!\! \!\! 
\Delta^2_p =  
\frac{{\sqrt s} }{  \Delta }  \, \Bigl [ (2n+1)  R_+  + (2m+1) R_-  \Bigr ]
\cr
\end{cases}
\label{delqp1}
\end{equation}
%
\medskip

\noindent
Concerning the regime $v<-1/2$, the calculation of
$\Delta^2_f =$ $ \langle \ell , k| W f^2W^+| k ,\ell\rangle$ is based on the more complex
unitary transformation $W = D_\phi S_\beta U_2 S_\alpha$ yielding
$$
\begin{cases}
&  
\!\!\!\!  
W x W^+ = \frac{e^\alpha}{\sqrt 2} 
\Bigl ( (c_\phi e^{-\beta} + s_\phi e^{\beta}) x + ( c_\phi e^{\beta} + s_\phi e^{-\beta}) q \Bigr ) 
\cr
& {\-} \cr
& 
\!\!\!\! 
W y W^+ = \frac{e^{-\alpha} }{\sqrt 2} 
\Bigl ( ( c_\phi e^{\beta} -s_\phi e^{-\beta} ) y+ ( c_\phi e^{-\beta} - s_\phi e^{\beta} )p \Bigr ) 
\cr
\end{cases}
,
$$
$$
\begin{cases}
&  
\!\!\!\!  
W q W^+ = \frac{e^{-\alpha} }{\sqrt 2} 
\Bigl ( ( c_\phi e^{\beta} -s_\phi e^{-\beta} )q - ( c_\phi e^{-\beta} -s_\phi e^{\beta}  ) x \Bigr ) 
\cr
& {\-} \cr
& 
\!\!\!\! 
W p W^+ = \frac{e^\alpha }{\sqrt 2} 
\Bigl ( ( c_\phi e^{-\beta} + s_\phi e^{\beta})p - ( c_\phi e^{\beta} + s_\phi e^{-\beta})y  \Bigr ) 
\cr
\end{cases}
,
$$
where one should recall that
$e^{2\alpha}= \sqrt {w}$ with $w =|v|-\frac{1}{2}$, 
$$
\eta = 2\sqrt 2 \tau /\sqrt w\, ,\,\, e^{2\beta}= [(\eta +2)/(\eta -2)]^{1/2}\, ,
$$ 
while $c_\phi = {\rm ch} (\phi)$ and $s_\phi = {\rm sh} (\phi)$. 
Deviations $\Delta^2_f$ can be shown to be function of
$$
{\rm ch} (2\phi) = {\eta (w+2)}/{D}\, ,\,\,\, {\rm sh} (2\phi) = -{w \sqrt{\eta^2-4} }/{D}\, ,
$$
with $D = 2\sqrt { \eta^2 (w+1) + w^2}$. As a consequence, deviations $\Delta_f^2$ can written
in terms of parameters $\eta$ and $w$
\begin{equation}
\begin{cases}
&  
\!\!\!\!  
\Delta^2_x = \frac{{\sqrt w}}{4} 
\left (  \frac{2k+1}{\nu_x^{2}} \chi_{-+} + \frac{2\ell+1}{\nu_q^{2}}  \chi_{+ +} \right )
\cr
& {\-} \cr
& 
\!\! \!\! 
\Delta^2_y = \frac{1}{4{\sqrt w}} 
\left ( (2k+1) \nu_x^2 \, \chi_{+-}  + (2\ell+1) \nu_q^2 \, \chi_{- -} \right )
\cr
\end{cases}
\label{delxy2}
\end{equation}

\begin{equation}
\begin{cases}
&  
\!\!\!\!  
\Delta^2_q = \frac{1}{4{\sqrt w}}  
\left ( \frac{2\ell+1}{\nu_q^{2}} \chi_{+-} + \frac{2k+1 }{\nu_x^{2}} \chi_{--} \right )
\cr
& {\-} \cr
& 
\!\! \!\! 
\Delta^2_p = \frac{{\sqrt w}}{4} 
\left ( (2\ell+1) \nu_q^2 \, \chi_{-+} + (2k+1) \nu_x^2 \, \chi_{+ +} \right )
\cr
\end{cases}
\label{delqp2}
\end{equation}
where $\chi_{rh} = ( e^{r\beta} + h e^{ -r\beta})$, $r= \pm$, $h = \pm$, while
$\nu^2_x $ and $\nu^2_q$ are defined by formulas (\ref{nuxq}).
%
%
The explicit form of symbols $\chi_{rh}$ is
$$
\chi_{-+} = \frac{2 {\cal R}^2_- }{D\sqrt {\eta^2-4}}\, ,\,\,\, 
\chi_{++} = \frac{2{\cal R}^2_+ }{D\sqrt {\eta^2-4}}\, ,
$$
with ${\cal R}_\pm = \sqrt {\eta^2+2w \pm D }$ and
$$
\chi_{+-} = 2\frac{w(\eta^2 -4) +{\cal R}^2_+}{D\sqrt {\eta^2-4}}\, ,\,\,\, 
\chi_{--} = 2\frac{w(\eta^2 -4) +{\cal R}^2_-}{D\sqrt {\eta^2-4}}\, .
$$
Note that, in the limit $\epsilon = \eta^2-4 \to 0$ one has
\begin{equation}
D \simeq 2(w+2) + \epsilon \frac{w+1}{w+2}\, ,\,\,\, 
{\cal R}_- \simeq \frac{\sqrt \epsilon}{\sqrt{w+2}}\, ,
\label{lim2}
\end{equation}
while ${\cal R}_+$ reduces to ${\cal R}_+ \simeq \sqrt{2(w+2)} $.

\bigskip


\end{document}